\documentclass[asa,12pt,psfig]{article}
\advance\topmargin by -0.75truein
\advance\textheight by 1.25truein
\def\mybibliography#1{{\noindent \Large \bf References}\list
 {}{\setlength{\leftmargin}{1em}\setlength{\labelsep}{0pt}
\itemindent=-\leftmargin}
 \def\newblock{\hskip .02em plus .20em minus -.07em}
 \sloppy\clubpenalty4000\widowpenalty4000
 \sfcode`\.=1000\relax}
\newbox\TempBox \newbox\TempBoxA
\def\uw#1{%
  \ifmmode\setbox\TempBox=\hbox{$#1$}\else\setbox\TempBox=\hbox{#1}\fi%
  \setbox\TempBoxA=\hbox to \wd\TempBox{\hss\char'176\hss}%
  \rlap{\copy\TempBox}\smash{\lower9pt\hbox{\copy\TempBoxA}}%
}
\newbox\TempBox \newbox\TempBoxA
\def\uwd#1{%
  \ifmmode\setbox\TempBox=\hbox{$#1$}\else\setbox\TempBox=\hbox{#1}\fi%
  \setbox\TempBoxA=\hbox to \wd\TempBox{\hss\char'176\hss}%
  \rlap{\copy\TempBox}\smash{\lower10pt\hbox{\copy\TempBoxA}}%
}
\def\mathunderaccent#1{\let\theaccent#1\mathpalette\putaccentunder}
\def\putaccentunder#1#2{\oalign{$#1#2$\crcr\hidewidth
\vbox to.2ex{\hbox{$#1\theaccent{}$}\vss}\hidewidth}}

\newcommand{\xtil}{\mathunderaccent\tilde{x}}

\newcommand{\betatil}{\mathunderaccent\tilde{\beta}}
\newcommand{\ffi}{\mathunderaccent\tilde{\phi}}

\usepackage{float}
\usepackage{color}
\usepackage{epsfig}
\usepackage{amsmath}
\usepackage{ntheorem}
\usepackage[T1]{fontenc}
\usepackage[latin1]{inputenc}
\usepackage[english]{babel}
\usepackage[font=small,labelfont=bf]{caption}
\usepackage{subfigure}
\usepackage{endnotes}
\usepackage{rotating}
\usepackage{bmpsize}
\usepackage{graphicx}
\usepackage{multirow}
\usepackage{wrapfig,lipsum,booktabs}
\numberwithin{equation}{section}
\newtheorem{theorem}{Theorem}
\newtheorem*{acknowledgement}[theorem]{Acknowledgement}

\usepackage{titlesec}

\titleformat{\section}{\normalfont\bfseries}{\thesection.}{0.5em}{}
 \titleformat{\subsection}{\normalfont\bfseries}{\thesubsection.}{0.5em}{}
 \titleformat{\subsubsection}{\normalfont\bfseries}{\thesubsubsection.}{0.5em}{}
\begin{document}

\vspace*{.10in}

\begin{center}

{\large \bf {Bayesian Inference of a Finite Population Mean\\ Under Length-Biased Sampling}}

\vspace*{.25in}

{  Zhiqing Xu,  Balgobin Nandram }\\
\vspace*{.25in}
Department of Mathematical Sciences, Worcester Polytechnic Institute\\
100 Institute Road, Worcester, MA 01609\\
(~zxu2@wpi.edu, balnan@wpi.edu)\\

\bigskip

{Binod Manandhar}\\
Department of Mathematics, University of Houston\\
(binod@math.uh.edu)
\medskip

November 17, 2018

\end{center}

\vspace*{.10in}

We present a robust Bayesian method to analyze forestry data when samples are selected
with probability proportional to length from a finite population of unknown size. Specifically, 
we use Bayesian predictive inference to estimate the finite population mean of shrub widths in a 
limestone quarry dominated by re-growth of mountain mahogany. The data on shrub widths are
collected using transect sampling and it is assumed that the probability that a shrub is selected
is proportional to its width; this is length-biased sampling. In this type of sampling, the
population size is also unknown and this creates an additional challenge. The quantity of interest
is average finite population shrub width and the total shrub area of the quarry can be estimated.
Our method is assisted by using the three-parameter generalized gamma distribution, thereby
robustifying our procedure against a possible model failure. Using conditional predictive 
ordinates, we show that the model, which accommodates length bias, performs better than the model
that does not. In the Bayesian computation, we overcome a technical problem associated with Gibbs 
sampling by using a random sampler.\\


\newpage

\section{Introduction}
Unequal probability sampling method was first suggested by Hansen and Hurwitz (1943) who 
demonstrated that the use of unequal selection probabilities frequently allowed more efficient 
estimators of the population total than did equal probability sampling. The sampling procedure Hansen and Hurwitz (1943) proposed was length-biased sampling. It occurs when the sample selection 
probabilities are correlated with the values of a study variable, e.g., size variable. This problem
falls under the general umbrella of selection bias problems in survey sampling.

Line intercept sampling is a length-biased method used to study certain quantitative characteristics 
of objects in a region of interest. In general, objects may be of any shape and size and may 
possess an arbitrary spatial distribution. For example, these objects may be shrubs or patches 
of vegetation in a field, or the projection of logs on the forest floor. The idea of line intercept 
sampling is to use lines (transects) as sampling units and measuring features of the objects (e.g., widths
of shrubs) that 
crossed by them. A length-biased sampling method producing samples from a weighted distribution. 
With the underlying distribution of population, one can estimate the attributes of the population 
by converting the weighted samples to random samples (surrogate samples). 

For the estimation of a finite population quantity, the problem is 
more complex than for a superpopulation parameter because if there is 
a bias which tends to make the sampled values large, the nonsampled 
values would tend to be small. Such an adjustment is difficult to 
carry out. Generally, it has been assumed that the sample size is much 
smaller than the population size, and this eliminates the finite 
population estimation problem. Recently Nandram, Bhatta and Bhadra (2013) 
proposed a Bayesian non-ignorable selection model to accommodate a 
selection mechanism for binary data; see also Nandram (2007). 

There are several approaches to address the selection bias problem.  One approach 
incorporates the nonsampled selection probabilities in a model. 
This approach is computer-intensive because 
the nonsampled part of the population is much larger than the sample;
e.g., see Nandram, Choi, Shen and Burgos (2006), Nandram and Choi (2010) 
and Choi, Nandram and Kim (2017).
The second approach involves two models, one for the sample, called 
the survey model, and the other for the population, called the 
census model. This approach is sometimes called the 
surrogate sampling approach; e.g., see Nandram (2007) and Nandram, Bhatta and
Bhadra (2013). The surrogate sampling approach obtains a surrogate random sample 
from the census model, and then prediction is done via the census model. 
The third approach is based on finite population sampling, in which a sample
distribution and a sample complement distribution are both constructed, see Sverchkov and Pfeffermann (2004). It is convenient to use this approach for line intersect sampling. 
Sverchkov and Pfeffermann (2004) developed design consistent predictors for the finite population 
total. Essentially they define the distributions of the sampled values 
and the nonsampled values as two separate weighted distributions of 
the census distribution (see Patil and Rao 1978). Yet another approach is based 
on quasi-likelihood (Chambers and Skinner 2003) which is difficult to perform in 
a Bayesian paradigm because the normalization constant is hard to evaluate (typically a complicated
function of the model parameters).
 
Length-biased distributions are a special case of the more general form known as weighted 
distributions. First introduced by Fisher (1934) to model ascertainment bias, weighted 
distributions were later formalized in a unifying theory by Rao (1965); see also the 
celebrated paper of Patil and Rao (1978). Briefly, if the random variable $x$ has a 
probability density function (pdf) of $f(x)$, and a non-negative weight function of $w(x)$, then the corresponding weighted density function is
$$g(x)=\frac{w(x)f(x)}{\int w(x)f(x) dx}.$$
A special case is when the weight function is $w(x)=x$. 
Such a distribution is known as a length-biased distribution and is given by
$$
g(x)=\frac{xf(x)}{\mu},
$$
where $\mu=\int xf(x)dx \ne 0$, subject to existence, is the expectation of $x$.
(In Bayesian statistics we do not use upper and lower cases to differentiate random variables and fixed 
quantities.)

Various works have been done to characterize relationships between the original distribution and the length-biased distribution.  Muttlak and McDonald (1990) suggested using a ranked set sampling procedure to estimate the population size and population mean. 
 
In this paper, we use a three-parameter generalized gamma distribution as the original distribution to model the widths of shrubs sampled by line intercept method. Line intercept method has been found in widespread applications when estimating particle density, coverage and yields. For example, Lucas and Seber (1977) and  Eberhardt (1978) derived unbiased estimators for density and percentage cover for any spatial distribution and randomly located transects. McDonald (1980) showed that the Lucas and Seber (1977) estimators for density and percentage cover are unbiased for a simple random sample of unequal length transects. Shrubs can be collected from either randomly located or systematically located transects (Butler and McDonald, 1983). It is evident that shrubs with larger widths have higher probabilities of selection.

In Section 2, we provide the background of our study. This includes data description and a introduction of the three-parameter generalized gamma distribution, which allows us to robustify our Bayesian
method to accommodate the length bias. Section 3 contains the model and the results. The procedure involves the following steps. First, we derive the population size distribution as well as the sample complement distribution by Bayes' theorem. Next, we propose a random sampling method to generate random parameters from their joint posterior distribution. Then, using each set of parameter values, we obtain a set of samples and the corresponding complement samples (Sverchkov and Pfeffermann, 2004). Finally, one sample of the finite population mean can be obtained by taking the average of the pooled samples. The goodness of fit is checked by utilizing conditional predictive ordinates, with results compared both for the models with and without the length bias. The conclusion is made in section 4.

\section{Data and Robustness}
\subsection{Description of the Data}
The data we use were collected using the line intercept sampling method (Muttlak and McDonald 1990). 
The study was conducted in a limestone quarry dominated by regrowth of mountain mahogany. The study area was defined by  the area east of the baseline and within the walls of the quarry, where the baseline was established approximately parallel to the fissures; see Figure 1 in Appendix A. By dividing the baseline into three equal parts, three systematically placed transects were established. To ensure uniform coverage over the study area, two independent replications, each with 3 transects were selected (see Figure 1). One quantity of 
interest is the mean width of the shrubs in the quarry. So the variable we study is the width of the 
projection of the shrub encountered by transects onto the baseline (for illustration see Figure 2). 

We use the data from both replications, as shown in Appendix A Tables 1 and 2. The numbers of shrubs counted in three transects are respectively $18, 22, 6$ and in the two transects of Replicate 2 are respectively $32, 11$ 
(one transect of Replicate 2 has no data). Looking at the box plots of these two replications (Figure 3), 
we notice the clear differences in the distributions among the three transects in Replication 1; whereas in Replication 2, the distributions are close. Therefore, when making inferences using Replication 1, we regard the data from these three transects as from three different strata (they are actually so), and distinguish them in our modeling.

One complication is that we do not know the number of shrubs in the entire quarry. As we intend to use 
the Bayesian approach, the data from Replicate 2 is used to construct a prior distribution of the finite population size and the data from Replicate 1 is to provide inference for the population mean.

\subsection{Generalized Gamma Distribution}
The generalized gamma distribution (GG) was first defined by Stacy (1962) and it encompasses various 
subfamilies including the Weibull distribution, the generalized normal distributions and the lognormal distribution
as a limit. Khodabin and Ahmadabadi (2010) provided details of the subfamilies of  
generalized gamma distribution. Because of the flexibility of generalized gamma distribution, we use
it as the underlying population distribution in our models. 

Some authors have advocated the use of simpler models because of estimation difficulties caused by the complexity of GG 
parameter structure. For example, Parr and Webster (1965), Hager and Bain (1971), and Lawless (1980) have considered maximum 
likelihood estimation in the three-parameter generalized gamma distribution. They reported problems with iterative solution 
of the nonlinear equations implied by the maximum likelihood method. They remarked that maximum likelihood estimators might 
not exist unless the sample size exceeds 400. (Our sample sizes are much smaller so we need to be careful.) In our paper, 
we perform Bayesian analyzes of generalized gamma distribution to overcome this issue.

The probability density of generalized gamma distribution is given by
\begin{equation}
f(x|\alpha,\beta,\gamma)=\frac{\gamma x^{\gamma\alpha-1}}{\beta^{\gamma\alpha}\Gamma(\alpha)}\exp\left[-\left(\frac{x}{\beta}\right)^\gamma\right],
x>0,
\end{equation}
where $\alpha$ $\beta$, $\gamma$ are all positive. It is worth noting that the
mean and variance of $x$ are given by
$$
E(x)=\beta \frac{\Gamma(\alpha+\frac{1}{\gamma})}{\Gamma(\alpha)},\quad\textrm{and}\quad
Var(x)= {\beta}^2 \left[\frac{\Gamma(\alpha+\frac{2}{\gamma})}{\Gamma(\alpha)}-
\left(\frac{\Gamma(\alpha+\frac{1}{\gamma})}{\Gamma(\alpha)}\right)^2\right],
$$
respectively. We write $x \sim GG(\alpha,\beta,\gamma)$ to denote a random variable with pdf, $f(x|\alpha,\beta,\gamma)$ defined by (2.1), which we call an unweighted generalized gamma distribution.

Note that when $\gamma=1$, we get the standard gamma distribution, and by making $\gamma$ differ
from $1$ many distributions are accommodated, thereby increasing the flexibility of the gamma distribution. It is in this sense we robustify our procedures.

%It is interesting to note that if we reparameterize the generalized gamma density using $\phi = \beta^{-\gamma}$,
%we get a simple form,
%$$
%f(x \mid \alpha, \phi, \gamma) = \frac{\gamma x^{\gamma\alpha-1}\phi^\alpha}{\Gamma(\alpha)}
%\exp\left[-\phi {x}^\gamma\right],
%x>0,
%$$
%a more convenient form for modeling and computation.

The length-biased distribution of sample x is $g(x)=\displaystyle \frac{xf(x)}{E(x)},$
where $E(x)$ is the expectation of $x$ from the unweighted density function $f(x)$. This can be easily derived as follows. Let $I$ denote the indicator variable, i.e., $I=1$ if the unit is selected and $I=0$ if the unit is not selected. Under length-biased sampling the probability that the unit has been selected given the value $x$ is $f(I=1|x) = Cx$, where $C$ is a constant. By Bayes' theorem, the sample pdf $g(x)$ is,
\begin{align}
g(x|I=1)
&=\frac{f(I=1|x)f(x)}{\int f(I=1|x)f(x)dx} \notag\\
&=\frac{Cx f(x)}{\int Cx f(x)dx} =\frac{x f(x)}{E(X)}. \notag
\end{align}
For convenience we will write $g(x)$ for $g(x|I=1)$.

Therefore, by using $GG(\alpha,\beta,\gamma)$ as the population distribution, the sample distribution is
\begin{align}
g(x|\alpha,\beta,\gamma)
&=\frac{\frac{\gamma x^{\gamma\alpha}}{\beta^{\gamma\alpha}\Gamma(\alpha)}\exp\left[-\left(\frac{x}{\beta}\right)^\gamma\right]}
{\beta\frac{\Gamma(\alpha+\frac{1}{\gamma})}{\Gamma(\alpha)}}\notag\\
&=\frac{\gamma x^{\gamma\alpha}}{\beta^{\gamma\alpha+1}\Gamma(\alpha+\frac{1}{\gamma})}\exp\left[-\left(\frac{x}{\beta}\right)^\gamma\right], x>0.
\end{align}
Note that $g(x)$ is also a generalized gamma distribution with parameters $\alpha_g=\alpha+\frac{1}{\gamma}, \beta_g=\beta$, and $\gamma_g=\gamma$, denote by $x \sim \mbox{GG}(\alpha+\frac{1}{\gamma}, \beta_, \gamma)$, with mean and variance adjusted to
$$
E(x)=\frac{\beta\Gamma(\alpha+\frac{2}{\gamma})}{\Gamma(\alpha+\frac{1}{\gamma})},\quad\textrm{and}\quad
Var(x)=\frac{{\beta}^2\Gamma(\alpha+\frac{3}{\gamma})}{\Gamma(\alpha+\frac{1}{\gamma})}-
\left(\frac{\beta\Gamma(\alpha+\frac{2}{\gamma})}{\Gamma(\alpha+\frac{1}{\gamma})}\right)^2.
$$
We call it the weighted generalized gamma distribution.

\section{Bayesian Methodology}
In this section, we derive the population size distribution,  the sample-complement distribution, as well as the posterior distribution for the parameters.

Denote $\ell$ as the number of transects, $N_i$ as the total number of shrubs in the $i^{th}$ transect,  $N=\sum_{i=1}^\ell N_i$. Note that all the $N_i$ and $N$ are unknown. Prior information about $N$ is needed to carry out a full Bayesian analysis.
Denote $n_i$ as the number of shrubs from the $i^{th}$ transect, $n = \sum_{i=1}^\ell n_i$ is the number of samples. Let $x_1,\ldots,x_n$ be the width of the sampled shrubs and $x_{n+1},\ldots,x_{N}$ be the widths  for the nonsampled ones, which are to be predicted. The quantity of interest is 
$$
\bar{X} = \frac{1}{N}\sum_{i=1}^N x_i = f \bar{x}_s + (1-f) \bar{X}_{ns},
$$
where $f=\frac{n}{N}$ is the sample fraction and $\bar{x}_s = \frac{1}{n}\sum_{i=1}^n x_i$ and $\bar{X}_{ns} = 
\frac{1}{N-n} \sum_{i=n+1}^N x_i$ are respectively 
the sample and nonsample means. A posteriori inference is required for  $\bar{X}_{ns}$. It is worth noting
that $\bar{X}_{ns}$ is not a sufficient statistic and cannot be derived directly from the sample. Therefore, one needs to draw $x_{n+1},\ldots,x_N$ to predict $\bar{X}_{ns}$.

In many studies the population size $N$ is unknown and it must be estimated before inference can be made about $\bar{X}$.
Our application is no exception, however, this is easy to address as we have two sets of replicated samples. The second replicate (samples from the two transects are similarly distributed) can be used to construct a prior for $N$. The first
replicate (three transects need to be treated as three strata) is used to estimate the population mean shrub width. In this way ``using the data twice'' is avoided.

We assume that the population distributions for different strata are GG,
$$
x_{ij} \mid \alpha,\beta_i,\gamma \stackrel{ind} \sim GG(\alpha, \beta_i, \gamma),  j=1,\ldots,N_i, i=1,\ldots,\ell,
$$
accommodated the length bias, the sample distribution,
\begin{equation}
x_{ij} \mid \alpha, \beta_i, \gamma \stackrel{ind} \sim GG(\alpha+\frac{1}{\gamma}, \beta_i, \gamma), j=1,\ldots,n_i, i=1,\ldots,\ell.
\end{equation}
The remaining problem is to find the distribution of the nonsampled values, $x_{ij},i=1,\ldots,\ell, j=n_i+1,\ldots,N_i$, the so called the sample complement distribution (Sverchkov and Pfeffermann, 2004), which we will describe later.

In  section 3.1 we show how to obtain the prior distribution for $N_i$. In section 3.2 we describe the sample complement distribution. In section 3.3, we combine the results of 3.1 and 3.2 to derive the full Bayesian Model. In section 3.4, we study the posterior distributions in detail. 

\subsection{Prior Distribution of the Finite Population Size}
We first find the estimate of $N$ based on the sample size. Then,  estimates of $N_i$ can be obtained assuming proportional allocation.

The Horvitz-Thompson unbiased estimator of $N$ is
$$\hat{N}=\sum\limits_{i=1}^n \frac{1}{\pi_i}$$
where $\pi_i$ is the probability that the $i^{th}$ unit is selected; see Cochran (1977). Since the 
line intercept sampling gives the length biased data, we are actually sampling with probability 
proportion to width $x$. Thus, we have
$$
\pi_i=Cx_i,\quad i=1,\dots,n,
$$
where $C$ is a constant and $C=\frac{1}{W}$, where $W=125$ (meters) is the length of the base line. Then, the 
estimated value of $N$ under selection bias is 
$$
\hat{N}=125\times \sum\limits_{i=1}^n \frac{1}{x_i},
$$
Using the data from the Replication 2, we have $\hat{N}=10,061$. (Note that Replication 1 has
$\ell = 2$ strata and Replication 2 has $\ell =3$ strata.)
Then, using proportional allocation in Replication 1, as $n_1=18$, $n_2=22$, $n_3=6$, we have $\hat{N}_1=3,937$, 
$\hat{N}_2=4,812$, $\hat{N}_3=1,312$. 

Next, we assume 
$$
n_i \mid N_i, \mu_o \stackrel{ind} \sim \mbox{Binomial}(N_i,\mu_o), n_i=0,\ldots,N_i, i=1,\ldots,\ell,
$$
 $\mu_o$ does not depend on transects because of the nature of proportional allocation method.
 
Using independent noninformative priors for $N_i$,
$$
\pi(N_i) \propto \frac{1}{N_i}, N_i \geq n_i.
$$
We derived the posterior distributions of $N_i$,
\begin{align}
\pi(N_i \mid n_i, \mu_o) = \frac{(N_i-1)!}{(n_i-1)!(N_i-n_i)!} \mu_0^{n_i} (1-\mu_0)^{N_i-n_i},
\quad N_i\ge n_i, i=1,\ldots,\ell,
\end{align}
which is a negative Binomial distribution with $E(N_i \mid n_i, \mu_o) = \displaystyle\frac{n_i}{\mu_o}$. By equating 
$\hat{N}_i$ to $E(N_i \mid n_i, \mu_o)$, we solve for the estimated value of $\mu_o$, which is  $\mu_o = \displaystyle \frac{n_i}{\hat{N}_i} = 0.0046$. 

Therefore, based on Replication 2 our data-based prior distributions of the $N_i$ are independently negative 
binomial distributions with parameters $n_i$ and $\mu_o=0.0046, i=1,\ldots,\ell$.

\subsection{Sample-Complement Distribution}
Next, we need to make inference about the non-sampled values. That is, we obtain the 
sample complement distribution (Sverchkov and Pfeffermann, 2004) and draw samples from it. We consider a single transect first and drop the transect indicator $i$.

Let $I_j=1$ if $j\in s$ and $I_j=0$ if $j\not\in s$, where $s$ denotes the sample set. Then
\begin{align*}
&I_j|x_j \sim Ber\left(\frac{x_j}{W}\right) ~\mbox{and}~ \; x_j \sim f(x_j)\\
&\Rightarrow \pi (I_j, x_j)\propto\left[\frac{x_j}{W} f(x_j) \right]^{I_j}\left[\left(1-\frac{x_j}{W}\right) f(x_j) \right]^{1-I_j}\\
&\Rightarrow \pi (x_j|I_j=0)=\frac{\left(1-\frac{x_j}{W}\right) f(x_j)}{\int \left(1-\frac{x_j}{W}\right) f(x_j)dx_j}.
\end{align*}

Thus, the posterior sample complement distribution given all the parameters is
\begin{align}
&\pi(x_{n+1},...,x_N|\alpha,\betatil ,\gamma,x_1,...,x_n,N) \notag\\
&=\prod\limits_{j=n+1}^N \frac{\left(1-\frac{x_j}{W}\right) f(x_j)}{\int \left(1-\frac{x_j}{W}\right) f(x_j) dx_j} = \prod\limits_{j=n+1}^N \left[ \frac{1-\frac{x_j}{W}}{1-\frac{\mu}{W}} \right] f(x_j),
\end{align}
where $f(x)$ is GG, $\mu$ is the expectation of $x$, which is $\mu=\frac{\beta\Gamma(\alpha+\frac{1}{\gamma})}{\Gamma(\alpha)}$.
We use the sampling importance re-sampling (SIR) algorithm to perform the sampling. The SIR algorithm is ideal because $\displaystyle\prod_{j=n+1}^N f(x_j)$ is a good proposal density and samples are easy to draw.

\subsection{Full Bayesian Model}
For the sample data our model is
$$
x_{ij} \mid \alpha, \beta_i, \gamma \stackrel{ind} \sim GG(\alpha+1/\gamma, \beta_i, \gamma),
j=1,\ldots,n_i
$$
and the priori for $\alpha$,  $\beta_i, i=1,\ldots,\ell$ and $\gamma$ are  
$$
\pi(\beta_i) \propto \frac{1}{\beta_i}, i=1,\ldots,\ell, \pi(\alpha) \propto \frac{1}{(1+\alpha)^2},
\pi(\gamma) \propto \frac{1}{(1+\gamma)^2}.
$$
Note that the priors on the $\beta_i$ are improper and the priors on $\alpha$ and $\gamma$ are the $f(2,2)$
distributions ($f(2,2)$ denotes the $f$ distribution with  degrees of freedom being $(2,2)$),
which are nearly noninformative (no moments exist) but proper.

The posterior sample complement distribution when incorporating all strata is
$$
\pi(x_{ij}, i=1,\ldots,\ell, j=n_i+1,\ldots\mid N_i ,\alpha, \betatil,\gamma)
= \prod_{i=1}^\ell \prod_{j=n_i+1}^{N_i} [\frac{1-\frac{x_{ij}}{W}}{1-\frac{\mu}{W}}] f(x_{ij} \mid \alpha, \betatil,\gamma),
$$ 
where $f$ and $\mu$ are defined in the same way as (3.3). 

The joint posterior density of $\alpha, \betatil, \gamma$ given
$\xtil_s = \{x_{ij}, j=1,\ldots,n_i, i=1,\ldots,\ell\}$ is
$$
\pi(\alpha, \betatil, \gamma \mid \xtil_s ) \propto \frac{\gamma^n \left(\prod\limits_{i=1}^{\ell} \prod\limits_{j=1}^{n_i} x_{ij}\right)^{\gamma \alpha}}{\left(\prod\limits_{i=1}^{\ell}\beta_i^{n_i}\right)^{\gamma \alpha+1}\left[\Gamma(\alpha+\frac{1}{\gamma})\right]^n}\exp\left[-\sum\limits_{i=1}^{\ell} \sum\limits_{j=1}^{n_i} {\left(\frac{x_{ij}}{\beta_i}\right)}^\gamma\right]\frac{1}{(1+\alpha)^2}  \frac{1}{(1+\gamma)^2 }\prod\limits_{i=1}^{\ell} \frac{1}{\beta_i} ,
$$

The posterior density can be simplified by transforming $\beta_i$ to $\phi_i=1/\beta_i^\gamma, i=1,\ldots,\ell$.
(Note that the jacobian of the transformation must be included.)
Then, the joint posterior density of $\alpha, \ffi, \gamma$ given
$\xtil_s$ is
\begin{equation}
\pi(\alpha,\ffi,\gamma \mid \xtil_s) \propto  \frac{1}{(1+\alpha)^2} \frac{1}{(1+\gamma)^2}
$$
$$
\times
\frac{\gamma^{n-\ell} \left(\prod\limits_{i=1}^\ell \prod\limits_{j=1}^{n_i} x_{ij}^{\gamma \alpha}  \right) \prod_{i=1}^\ell \phi_i^{n_i\left(\alpha+\frac{1}{\gamma}\right)-1}}{\left[\Gamma(\alpha+\frac{1}{\gamma})\right]^n}
\exp\left[-\sum\limits_{i=1}^\ell\phi_i \sum\limits_{j=1}^{n_i}  x_{ij}^\gamma\right].
\end{equation}
This is not a standard posterior density, however, one can fit this model using Markov chain
Monte Carlo methods (i.e., to obtain sample of $\alpha, \ffi, \gamma$).

\subsection{Further Study of the Posterior Density}
One important problem we need to worry about is the posterior propriety of 
$\pi(\alpha,\ffi,\gamma \mid \xtil_s)$. 
First, it is easy to see that
\begin{equation}
\label{gam}
\phi_i \mid \alpha, \gamma, \xtil_s \stackrel{ind} \sim
\mbox{Gamma}\{n_i(\alpha+\frac{1}{\gamma}),  \sum_{j=1}^{n_i} x_{ij}^\gamma\}, i=1,\ldots,\ell.
\end{equation}
Then,  integrating out
the $\phi_i$, we get
\begin{equation}
\label{ipos}
\pi(\alpha,\gamma \mid \xtil_s) \propto 
\prod_{i=1}^\ell 
\left\{ \gamma^{n_i-1}
\frac{ \left(\prod\limits_{j=1}^{n_i} x_{ij}^{\gamma \alpha} \right)}
{\left(\sum\limits_{j=1}^{n_i}  x_{ij}^\gamma \right)^{n_i(\alpha+\frac{1}{\gamma})}}
\frac{\Gamma\{n_i\left(\alpha+\frac{1}{\gamma}\right)\}}
{\left(\Gamma(\alpha+\frac{1}{\gamma})\right)^{n_i}}
\right\}
\frac{1}{(1+\alpha)^2} \frac{1}{(1+\gamma)^2}.
\end{equation}

It is convenient to let $a_i = \sum_{j=1}^{n_i} x_{ij}^\gamma/n_i$ and $g_i = \left(\prod_{j=1}^{n_i} x_{ij}^\gamma\right)^{1/n_i}$
denote the arithmetic and geometric means of the $x_{ij}^{\gamma}, j=1,\ldots,n_i, i=1,\ldots,\ell$. Then,
we have
\begin{equation}
\label{spos}
\pi(\alpha,\gamma \mid \xtil_s) \propto 
\prod_{i=1}^\ell 
\left\{ \left(\frac{g_i}{a_i}\right)^{n_i\alpha} \left(\frac{\gamma^{n_i-1}}{a_i^{n_i/\gamma}} \right)
\frac{\Gamma\{n_i\left(\alpha+\frac{1}{\gamma}\right)\}}
{n_i^{n_i(\alpha+\frac{1}{\gamma})}\left(\Gamma(\alpha+\frac{1}{\gamma})\right)^{n_i}}
\right\}
\frac{1}{(1+\alpha)^2} \frac{1}{(1+\gamma)^2}.
\end{equation}

Thus, we essentially have a two-parameter posterior density. Although an overkill, we attempted to fit this model 
using  a Gibbs sampler. There are  difficulties in performing the Gibbs sampler (perhaps associated with the difficulties encountered in finding MLEs in generalized gamma distribution) because high correlations are 
present among the parameters and thinning is not helpful. The problem is essentially high correlations between
$\alpha$ and $\gamma$. Thus, we consider an alternative algorithm which simply uses the multiplication rule
of probability. We prove the theorem below which adds credence to our Bayesian methodology.

However, since $\gamma=1$ makes the generalized gamma density a standard gamma density, it is sensible to bound
$\gamma$ in an interval centered at $1$. That is, we take  $a_o^{-1} \leq \gamma \leq a_o$; a sensible choice is
$a_0 = 10$ or so. Thus,  we replace the prior on $\gamma$ by $\gamma \sim \mbox{Uniform}(a_o^{-1}, ~a_o)$; the
original prior is inconvenient and not helpful.

{\bf Theorem}

Assuming that  $a_o^{-1} \leq \gamma \leq a_o$, the joint posterior density of $\pi(\alpha, \gamma \mid \xtil_s)$ 
is proper.

{\bf Remark}: Using the multiplication rule of probability,
$$
\pi(\ffi,\alpha,\gamma \mid \xtil_s) = \pi(\ffi \mid \alpha,\gamma, \xtil_s) 
\pi(\alpha,\gamma \mid \xtil_s),
$$
the theorem implies that $\pi(\ffi,\alpha,\gamma \mid \xtil_s)$ is also proper. Clearly,
$\pi(\betatil,\alpha,\gamma \mid \xtil_s)$ is also proper.

{\bf Proof}

We make two observations.
First, using the arithmetic-geometric inequality, we have
$\left(\frac{g_i}{a_i}\right)^{n_i\alpha}  \leq 1$.
Second, using $a_o^{-1} \leq \gamma \leq a_o$, we have 
$\frac{\gamma^{n_i-1}}{a_i^{n_i/\gamma}}$, a function of $\gamma$, is bounded uniformly in $\gamma$.
Therefore, we only need to show that
$$ I =
\int_{a_o^{-1}}^{a_o} \int_{0}^\infty
\left\{ \prod_{i=1}^\ell
\frac{\Gamma\{n_i\left(\alpha+\frac{1}{\gamma}\right)\}}
{n_i^{n_i(\alpha+\frac{1}{\gamma})}\{\Gamma(\alpha+\frac{1}{\gamma})\}^{n_i}} \right\}
\frac{1}{(1+\alpha)^2} \frac{1}{a_o-a_o^{-1}}
d\alpha d\gamma < \infty.
$$
%
%$$
%I = 
%\int_{0}^{\infty} \int_{0}^\infty
%\left\{ \prod_{i=1}^\ell
%\frac{\Gamma\{n_i\left(\alpha+\frac{1}{\gamma}\right)\}}
%{n_i^{n_i(\alpha+\frac{1}{\gamma})}\{\Gamma(\alpha+\frac{1}{\gamma})\}^{n_i}} \right\}
%\frac{1}{(1+\alpha)^2} \frac{1}{(1+\gamma)^2} 
%d\alpha d\gamma < \infty.
%$$

Next, we transform $\alpha$ to $\theta = \alpha+\frac{1}{\gamma}$, keeping $\gamma$ untransformed.
Then, the integral becomes
$$ I = 
\int_{a_o^{-1}}^{a_o} \int_{\frac{1}{\gamma}}^\infty
g^\ast(\theta)
\frac{1}{(1+\theta-\frac{1}{\gamma})^2} \frac{1}{a_o-a_o^{-1}}
d\theta d\gamma < \infty,
$$
where
$$
g^\ast(\theta) = 
\prod_{i=1}^\ell
\frac{\Gamma(n_i \theta)}
{n_i^{n_i\theta}\{\Gamma(\theta)\}^{n_i}}.
$$
%
%Next, we transform $\alpha$ to $\theta = \alpha+\frac{1}{\gamma}$, keeping $\gamma$ untransformed.
%Then, the integral becomes
%$$
%I = 
%\int_{0}^{\infty} \int_{\frac{1}{\gamma}}^\infty
%g^\ast(\theta)
%\frac{1}{(1+\theta-\frac{1}{\gamma})^2} \frac{1}{(1+\gamma)^2} 
%d\theta d\gamma < \infty,
%$$
%where
%$$
%g^\ast(\theta) = 
%\prod_{i=1}^\ell
%\frac{\Gamma(n_i \theta)}
%{n_i^{n_i\theta}\{\Gamma(\theta)\}^{n_i}}.
%$$
We only need to show that 
$$
g_i(\theta) = 
\frac{\Gamma(n_i \theta)}
{n_i^{n_i\theta}\{\Gamma(\theta)\}^{n_i}}
$$
is bounded uniformly in $\theta$ for any $i=1,\ldots,\ell$. For convenience, we will drop the subscript, $i$,
momentarily, so we simply need to show that $\Delta(\theta) = \tilde{\Gamma}(n\theta) - n\tilde{\Gamma}(\theta) - n \theta\ln(n)$, where $\tilde{\Gamma}(\cdot)$ is the
logarithm of the gamma function, is uniformly bounded in $\theta$; see the Appendix A. 

Finally, because $g^\ast(\theta) \leq A < \infty$, we are left with 
$$
I \leq
A\int_{a_o^{-1}}^{a_o} \frac{1}{(a_o-a_o^{-1})}  \left\{\int_{\frac{1}{\gamma}}^\infty
\frac{1}{(1+\theta-\frac{1}{\gamma})^2} 
d\theta\right\} d\gamma = A\int_{a_o^{-1}}^{a_o} \frac{1}{(a_o-a_o^{-1})} d\gamma  \int_{0}^\infty
\frac{1}{(1+\alpha)^2} 
d\alpha = A.
$$
Therefore, our claim on propriety holds.

\section{Bayesian Computations and Data Analyzes}
In this section, we perform Bayesian analysis of the posterior distributions 
of  population parameters by a numerical method, called random sampler, which performs better than the Gibbs sampler .
We then obtain the nonsampled values using the sampling importance resampling (SIR) algorithm. Recall that the data we use here is 
Replication I, which has three transects, i.e., $\ell = 3$.

\subsection{Random Sampler}
Since the Gibbs sampler is a Markovian updating scheme, we have shown, in our work not presented in this paper, that most of the estimated values of population mean are larger than what we expected (see results in Appendix Table 3). One of the reasons is that high correlations among these parameters makes the Gibbs sampler inefficient in the sense it may take a very large number of iterations to converge in distribution. In this section, we propose a non-Markovian algorithm, called random sampler, in order to avoid the particular issue mentioned above.

Therefore, $\alpha$ and $\gamma$ cannot be sampled directly from their unbounded parameters space. We use the transformation $\alpha ' =\frac{\alpha}{1+\alpha}$ and $\gamma ' =\frac{\gamma}{1+\gamma}$.
Then,
\begin{align*}
&\Pi(\alpha ', \gamma '|x_{11}, \cdots, x_{3n_3})=\int_{\phi_1} \int_{\phi_2} \int_{\phi_3} \Pi(\alpha,\phi_1,\phi_2,\phi_3,\gamma|x_{11}\cdots, x_{3n_3}) d \phi_1 d\phi_2 d\phi_3 \\
&=\left \{\frac{\gamma^n \left(\prod\limits_{i=1}^3 \prod\limits_{j=1}^{n_i} x_{ij}\right)^{\gamma \alpha}}{\left[\Gamma(\alpha+\frac{1}{\gamma})\right]^n} \frac{\Gamma\left(n_1(\alpha+\frac{1}{\gamma})\right) \Gamma\left(n_2(\alpha+\frac{1}{\gamma})\right) \Gamma\left(n_3(\alpha+\frac{1}{\gamma})\right) }{\left(\sum x_{1j}^{\gamma}\right)^{n_1 (\alpha+\frac{1}{\gamma})}
\left(\sum x_{2j}^{\gamma}\right)^{n_2 (\alpha+\frac{1}{\gamma})}
\left(\sum x_{3j}^{\gamma}\right)^{n_3 (\alpha+\frac{1}{\gamma})} {\gamma} ^3}\right \}_{\alpha=\frac{\alpha '}{1-\alpha '}, \gamma=\frac{\gamma'}{1-\gamma'}}, \\
&\alpha ' \in (0, 1), \gamma ' \in (0, 1) .
\end{align*}

Two-dimensional grid method can be applied to draw $\alpha '$ and $\gamma '$ from their joint distribution. But grid method is computationally intensive in more than one dimension. We used the multiplication rule to draw samples of $\alpha '$ and $\gamma '$.
\begin{align}
\Pi(\alpha ', \gamma '| x_{11}\cdots, x_{3n_3})=\Pi(\alpha '| \gamma ', x_{11}\cdots, x_{3n_3}) \Pi(\gamma '| x_{11}\cdots, x_{3n_3}) .
\end{align}

To apply this rule, we fist generated a sample of ${\gamma '}^{(1)}$ from $\Pi(\gamma '| x_{11}\cdots, x_{3n_3})$, then generated a sample of ${\alpha '}^{(1)}$ from $\Pi(\alpha '| {\gamma '}^{(1)}, x_{11}\cdots, x_{3n_3})$. Repeating this procedure $M$ times to obtain $M$ sets of $\alpha ' (\alpha)$ and $\gamma ' (\gamma)$. The corresponding $\phi (\beta)$ can also be obtained by sampling from $\Pi(\phi_i|\alpha,\phi_k, \gamma, x_{11}, \cdots, x_{3n_3})$.

The term $\Pi(\alpha '| \gamma ', x_{11}\cdots, x_{3n_3})$ in (4.1) is easy to derive.
\begin{align*}
&\Pi(\alpha '| \gamma ', x_{11}\cdots, x_{3n_3}) \\
&\propto \left \{\frac{\gamma^n \left(\prod\limits_{i=1}^3 \prod\limits_{j=1}^{n_i} x_{ij}\right)^{\gamma \alpha}}{\left[\Gamma(\alpha+\frac{1}{\gamma})\right]^n} \frac{\Gamma\left(n_1(\alpha+\frac{1}{\gamma})\right) \Gamma\left(n_2(\alpha+\frac{1}{\gamma})\right) \Gamma\left(n_3(\alpha+\frac{1}{\gamma})\right) }{\left(\sum x_{1j}^{\gamma}\right)^{n_1 (\alpha+\frac{1}{\gamma})}
\left(\sum x_{2j}^{\gamma}\right)^{n_2 (\alpha+\frac{1}{\gamma})}
\left(\sum x_{3j}^{\gamma}\right)^{n_3 (\alpha+\frac{1}{\gamma})}}\right \}_{\alpha=\frac{\alpha '}{1-\alpha '}} ,\\
&\alpha ' \in (0,1) .
\end{align*}

The term $\Pi(\gamma '| x_{11}\cdots, x_{3n_3})$ in (4.1) can be derived by integrating $\Pi(\alpha ', \gamma '| x_{11}\cdots, x_{3n_3})$ with respect to $\alpha '$. Unfortunately, it is not possible to integrate $\Pi(\alpha ', \gamma '| x_{11}\cdots, x_{3n_3})$ by analytical techniques. For this reason, numerical methods have to be used. We use the 20-point Gaussian quadrature to approximate $\Pi(\gamma '| x_{11}\cdots, x_{3n_3})$.
\begin{align*}
\Pi(\gamma '| x_{11}\cdots, x_{3n_3}) = &\int_0^1 \Pi(\alpha ', \gamma '| x_{11}\cdots, x_{3n_3}) d\alpha ' \\
&=\frac{1}{2} \int_{-1}^1 \Pi \left(\frac{1}{2} + \frac{1}{2}\alpha ', \gamma ' \right) d\alpha ' \\
&\approx \frac{1}{2} \sum_{i=1}^{20} \omega_i \Pi\left(\frac{1}{2} + \frac{1}{2} x_i , \gamma ' \right),
\end{align*}
where $x_i, i=1,\cdots, 20$ are the roots of orthogonal polynomials $P_{20}(x)$ for $[-1,1]$ and $\omega_i, i=1, \cdots ,20$ are the corresponding Gauss-Legender weights, which can be created by R package 'gaussquad'.

The summary of samples drawn by Random Sampler for each parameter and the population mean are shown in Table 6 . It is shown that the population mean has the IQR of $(.67, .81)$, with the median of $.75$.
In Fig. 4, it is shown that the posterior distribution of $\alpha$ is bimodal (pointing to the difficulty in estimating $\alpha$), while those of $\beta_1$, $\beta_2$, $\beta_3$ and $\gamma$ are skewed to the right.
In the next section, we will perform the model checking by conditional predictive ordinate (CPO).

\subsection{Model Checking by Conditional Predictive Ordinate}
Comparing the predictive distribution to the observed data is generally termed a ``posterior predictive check''. This type of check includes the uncertainty associated with the estimated parameters of the model. Posterior predictive checks (via the predictive distribution) involve a double-use of the data, which causes predictive performance to be overestimated. To overcome this drawback, Geisser and Eddy (1979) has proposed the leave-one-out cross-validation predictive density. This is also known as the conditional predictive ordinate or CPO (Gelfand, 1996).
The CPO is a handy posterior predictive check because it may be used to identify outliers, influential observations, and for hypothesis testing across different non-nested models. The CPO expresses the posterior probability of observing the value of $x_i$ when the model is fitted to all data except $x_i$, with a larger value implying a better fit of the model to $x_i$, and very low CPO values suggest that $x_i$ is an outlier and an influential observation.
A Monte Carlo estimate of the CPO is obtained without actually omitting $x_i$ from the estimation, and is provided by the harmonic mean of the likelihood for $x_i$. Specifically, the CPO$_i$ is the inverse of the posterior mean of the inverse likelihood of $x_i$.
The Monte Carlo estimate of CPO is
\begin{align*}
&\widehat{CPO_i}=\left[\frac{1}{M} \sum \limits_{h=1}^M \frac{1}{f(x_i|\tilde{\theta}^{(h)})}\right]^{-1}, i
=1,2,...,n,
\end{align*}
where $\tilde{\theta}^{(h)}\overset{iid}{\sim} \Pi(\tilde{\theta}|\tilde{x})$ for $h=1,\dots, M$;
see Molina, Nandram and Rao (2014).

The sum of the log(CPO)'s can be an estimator for the natural logarithm of the marginal likelihood, sometimes called the log pseudo marginal likelihood (LPML)
$$
LPML=\sum \limits_{i=1}^n log(\widehat{CPO_i}).
$$
Models with larger LPMLs are better.
To compare the predictive distributions (both model with length bias and model without length bias) using our length-biased sample, we calculated the LPML for both models.
The likelihood of $x_i$ under length biased model is given by
\begin{align*}
f(x_i|\alpha,\beta,\gamma)=\frac{\gamma x_i^{\gamma\alpha}}{\beta^{\gamma\alpha+1}\Gamma(\alpha+\frac{1}{\gamma})}\exp\left[-\left(\frac{x_i}{\beta}\right)^\gamma\right] ,
\end{align*}
where $\beta$ is the corresponding parameter for the stratum that $x_i$ is from.
The likelihood of $x_i$ under no length biased model is given by
\begin{align*}
f(x_i|\alpha,\beta,\gamma)=\frac{\gamma x_i^{\gamma\alpha-1}}{\beta^{\gamma\alpha}\Gamma(\alpha)}exp\left[-\left(\frac{x_i}{\beta}\right)^\gamma\right],
\end{align*}
where $\beta$ is the corresponding parameter for the stratum that $x_i$ is from.
It is shown that the $LPML$ of for model with length bias is larger (LPML =$ -36.10$) than the one for
the model without length bias (LPML =$ -47.54$), which means the model with length bias fits our length-biased sample better.

\subsection{Nonsampled Widths}
We define the importance function as
\begin{equation}
\Pi_a(x_{n+1},\cdots, x_N|N)=\frac{\prod \limits_{i=n+1}^N f(x_i)}{ \displaystyle \int \prod \limits_{i=n+1}^N f(x_i) dx_{n+1} \cdots dx_N} .
\end{equation}
Then, the importance ratios are
\begin{align}
\frac{\Pi(x_{n+1},\cdots, x_N|N)}{\Pi_a(x_{n+1},\cdots, x_N|N)} \propto \prod \limits_{i=n+1}^N \frac{1-\frac{x_i}{W}}{1-\frac{\mu}{W}} .
\end{align}
A random sample can now be obtained by re-sampling with probability proportional to the ratios.

The algorithm to obtain the nonsampled values is as follows.
\begin{itemize}
\item Step 1. Obtain M sets of $(\alpha, \beta_1,\beta_2,\beta_3,\gamma,)$ using the sampling methods described in the next Section.
\item Step 2. Obtain a sample of $N$ from formula (3.2).
\item Step 3. For each set of parameters, generate the vector $\tilde{x_j}$ where $x_{ij}, i=n_j+1, \cdots, (N_j-n_j), \; j=1,2,3$ from the corresponding generalized gamma distribution.
\item Step 4. Compute the population mean and the importance ratio $w$.
\item Step 5. Repeat the step 2 to step 4 $M-1$ times.
\item Step 6. Draw $\alpha M$ values of the population mean with probability proportional to $\tilde{w}$. We choose $\alpha=.1$
\end{itemize}

\section{Summary}
In this paper we have presented a model for estimating population mean under length-biased sampling. We have used a weighted distribution of the three-parameter generalized gamma distribution to model the shrub widths to robustify our procedure that accommodates the length-biased sampling. Interest is  on the finite population mean of shrub width in the entire quarry. In order to avoid certain technical issues associated with classical inference when using the generalized gamma distribution, we proposed a non-Markovian Bayesian numerical method, called random sampler, which performs better than Gibbs sampler when the population parameters are highly correlated. Posterior population distributions are easily estimated using this method. Conditional predictive ordinate shows that the model with length bias performs better than the model without length bias. 

While accommodating transect sampling is a challenge, another important challenge in our procedure is to estimate the unknown population size. To ensure a full Bayesian procedure,
we have used the data in Replication 2 (two strata) to estimate the finite population size that is usually unknown in this type of problem. The data from Replication 1 were used to estimate the average shrub width of the
finite population. This estimate can, in turn, be used to give an estimate of the total shrub area assuming a
standard geometry (e.g., a circle with the width being the diameter or a square with the width being length
of a side). 

An interesting topic for future research would be including covariates to study potential predictors.  In Muttlak and McDonald (1990), in addition to the measurement of shrub widths, two more attributes of mountain mahogany, maximum height, and number of stems, were measured. Both attributes are important predictors of the average shrub width of an area's vegetation. Semi-parametric linear regression (Chen, 2010) or generalized linear regression can be considered to measure this association.
We can incorporate the covariates through a gamma type regression model. Let the covariates be $\underset{\sim}{z_{ij}}, i=1,2,3, j=n_1, n_2, n_3$ and $\underset{\sim}{\phi_i}, i=1,2,3$.
Because the mean of each stratum is linearly related to $\beta_1, \beta_2, \beta_3$ respectively, we take $\beta_i=\displaystyle e^{\underset{\sim}{z_i} ' \underset{\sim}{\phi_i}} , i=1,2,3$.
For the shrub data,  our model is $$
P(\underset{\sim}{x}|\underset{\sim}{z},\underset{\sim}{\phi}, \alpha, \gamma)
=\prod_{i=1}^{3} \prod_{j=1}^{n_i} \frac{\gamma x_{ij} ^{\gamma \alpha -1} \left[e^{-z_{ij} \phi_i}\right] ^{\gamma \alpha}}{\Gamma(\alpha)} \exp{\{-\left(x_{ij} e^ {-z_{ij} \phi_i}\right)^\gamma\}} .
$$
A similar form can be easily written down for the length-biased sampling. Our future plan is to fit a model to accommodate the covariates.

\medskip

\begin{acknowledgement}
The author thanks the Associate Editor, XXX, and the referee for their comments which improved the quality of the paper very much.        
\end{acknowledgement}
\medskip

\appendix

\begin{flushleft}
\noindent \normalsize\bf{Appendix A. Uniform Boundedness of $\Delta(\theta)$}\\
\end{flushleft}
%%%%%%%%%%%%%%%%%%%%%%%%%%%%%%%%%%%%%%%%%%%%%%%%%%%%%%%

We need to show that
$$
\Delta(\theta) = \tilde{\Gamma}(n\theta) - n\tilde{\Gamma}(\theta) - n \theta\ln(n)
$$
is uniformly bounded in $\theta$. We will show that $\Delta(\theta)$ asymptotes out horizontally.

First, differentiating $\Delta(\theta)$, we have
$$
\Delta^\prime(\theta) = n\{\psi(n\theta)-\psi(\theta)-\ln(n)\},
$$
where $\psi(\cdot)$ is the diagamma function.  Now, using the duplication property (Abramowitz and Stegun 1965, Ch. 6) 
of the digamma function, one can show that $\Delta^\prime(\theta) \geq 0$.
That is, $\Delta(\theta)$ is monotonically increasing in $\theta$; see also Figure 5.

Second, differentiating $\Delta^\prime(\theta)$, we have
$$
\Delta^{\prime\prime}(\theta) = \frac{n}{\theta} \{n\theta \psi^\prime(n\theta)  - \theta \psi^\prime(\theta)\}.
$$
Using a theorem (Ronning 1986) which states that $x\psi^\prime(x)$ decreases monotonically in $x$, we have
$\Delta^{\prime\prime}(\theta) \leq 0$. That is, $\Delta(\theta)$ is concave and the rate of increase of 
$\Delta(\theta)$ decreases; see Figure 6.

Therefore, $\Delta(\theta)$ asymptotes out horizontally and $\Delta(\theta)$ must 
be bounded; so is its exponent.

\newpage
\begin{table}[htbp]
	\centering
	\begin{tabular}{cc}
		\hline
		Transect & $X_i$=width\\ \hline
		I & 1.53, 0.87, 0.79, 0.78, 1.85, 1.45 \\&0.48, 0.52, 0.22, 0.38, 0.59, 0.20\\& 0.42, 1.02, 0.97, 0.56, 0.62, 0.42\\ \hline
		II & 1.15, 0.87, 0.57, 0.97, 0.57, 1.97\\& 0.58, 2.54, 1.85, 0.35, 1.24, 1.80 \\&0.78, 0.98, 1.30, 1.55, 1.69, 2.12\\& 1.27, 0.75, 1.01, 1.82 \\ \hline
		III & 0.71, 1.50, 1.82, 1.86, 1.61, 1.21\\ \hline
	\end{tabular}
	\caption{Widths (meters) of shrubs in Replication 1}
\end{table}

\begin{table}[htbp]
	\centering
	\begin{tabular}{cc}
		\hline
		Transect & $X_i$=width\\ \hline
		I & 0.67, 0.31, 0.83, 1.95, 1.36, 1.45 \\&0.72, 1.15, 0.98, 1.29, 0.88, 0.25 \\&0.63, 1.12, 0.34, 0.21, 1.36, 0.95\\& 1.04, 0.48, 1.05, 0.88, 0.16, 1.08\\& 0.95, 0.25, 0.30, 1.40, 0.58, 0.73\\& 1.30, 0.57\\ \hline
		II &  0.96, 2.08, 0.68, 1.39, 0.50, 0.72 \\&0.19, 1.91, 0.88, 0.48, 0.12\\ \hline
	\end{tabular}
	\caption{Widths (meters) of shrubs in Replication 2}
\end{table}

\newpage

\begin{table}[htbp]
	\centering
	\renewcommand{\arraystretch}{1.5}
	\begin{tabular}{cccccc}
		\hline
		$\alpha$ & $\beta_1$ & $\beta_2$ & $\beta_3$ & $\gamma$ & $\bar{x}$\\ \hline
		1.17&0.71&1.41&1.24&1.7&0.93\\ \hline
	\end{tabular}
	\caption{Estimated parameters and population mean in the model  with corrected selection bias. 95\% Bootstrap Confidence Intervals of $\bar{x}$ is (.86, 1.11).}
\end{table}

\begin{table}[htbp]
	\centering
	\renewcommand{\arraystretch}{1.5}
	\begin{tabular}{cccccc}
		\hline
		$\alpha$ & $\beta_1$ & $\beta_2$ & $\beta_3$ & $\gamma$ & $\bar{x}$\\ \hline
		1.80&0.71&1.41&1.25&1.54&1.46\\ \hline
	\end{tabular}
	\caption{Estimated parameters and population mean in the model with uncorrected selection bias. 95\% Bootstrap Confidence Intervals of $\bar{x}$ is (1.38, 1.72).}
\end{table}

\newpage
\begin{table}[htbp]
	\centering
	\begin{tabular}{ccccccc}
		\hline
		Name& Min. & 1st Qu. & Median & Mean & 3rd Qu. & Max.\\ \hline
		$\alpha$ & 0.25 & 0.28 &1.62& 2.86 &5.36&6.78 \\
		$\beta_1$ &0.25&  0.25&  0.40&  0.55&  0.84&  1.25\\
		$\beta_2$ &0.25&  0.25&  0.62&  0.82&  1.33&  1.91\\ 
		$\beta_3$ &0.25&  0.25&  0.53&  0.92&  1.33&  4.83\\
		$\gamma$ &0.48&  0.58&  0.83&  1.06&  1.35&  2.50\\
		$\bar{X}$ &0.066&  0.088&  3.44& 13.92&26.01& 58.86\\ \hline
	\end{tabular}
	\caption{Summary of the Parameters and Population Mean by Gibbs Sampler}
\end{table}

\begin{table}[htbp]
	\centering
	\begin{tabular}{ccccccc}
		\hline
		Name& Min. & 1st Qu. & Median & Mean & 3rd Qu. & Max.\\ \hline
		$\alpha$ &0.25&  0.77&  1.37&  1.34&  1.93&  2.33\\
		$\beta_1$  &0.07&  0.29&  0.46&  0.52&  0.71&  1.50\\
		$\beta_2$ &0.17&  0.49&  0.75&  0.83&  1.12&  2.19\\
		$\beta_3$ &0.13&  0.56&  0.87&  0.96&  1.29&  2.98\\
		$\gamma$ &0.64&  1.05&  1.36&  1.43&  1.68&  3.54\\
		$\bar{X}$ & 0.31 & 0.67 &0.75& 0.74 &0.81&1.01\\ \hline
	\end{tabular}
	\caption{Summary of the Parameters and Population Mean by Random Sampler}
\end{table}

\begin{figure}[H]
  \begin{center}
\includegraphics[height=5.5in]{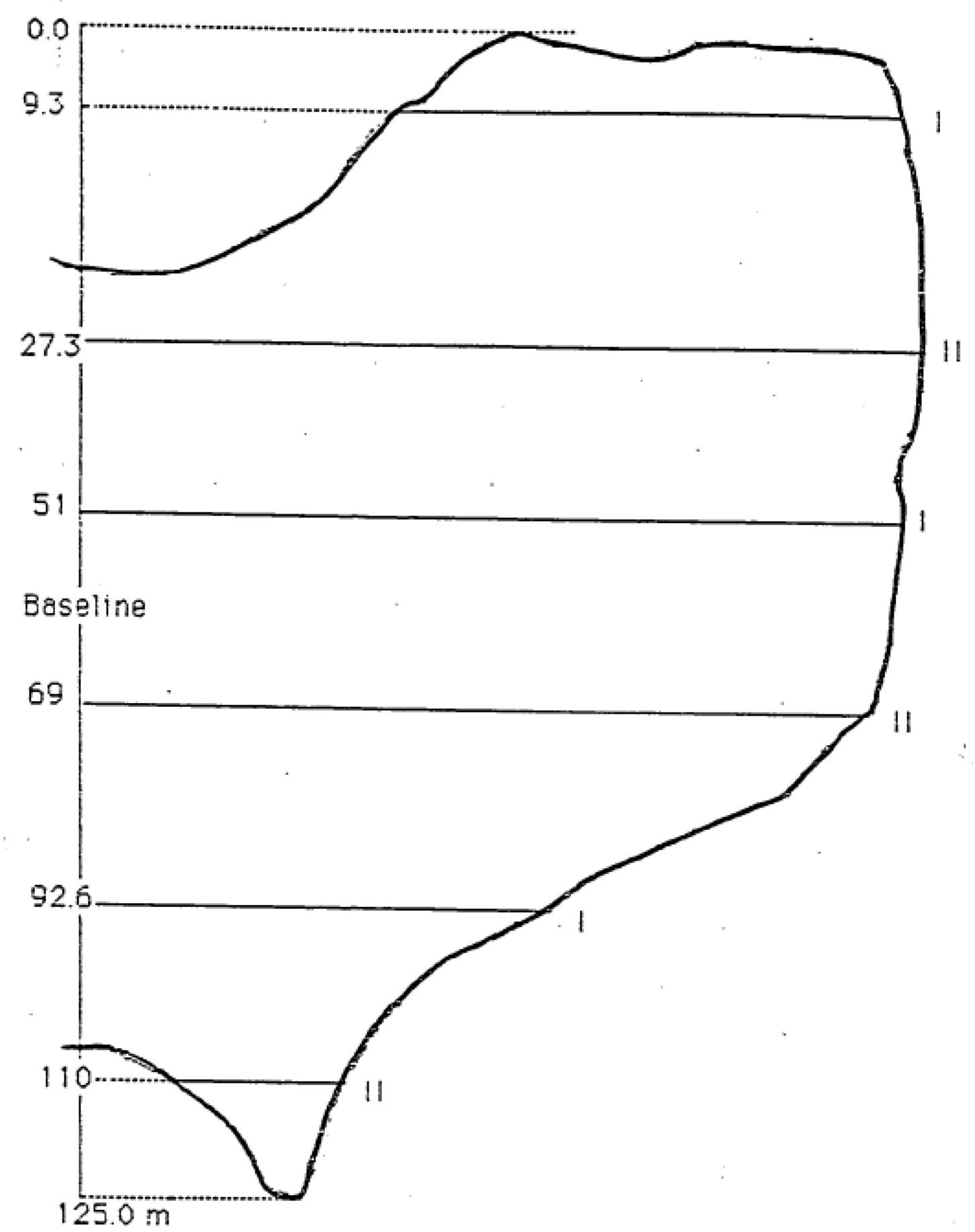}
  \end{center}
  \caption{Sketch of the study area showing the baseline and transects of the two replicates (I and II)
  of systematically located transect lines; Random starting points of 9.3 m and 27.3 m were selected
  with parallel lines separated by 41.66 m; see Muttlak (1988)}
\end{figure}

\newpage
 \begin{figure}[H]
  \begin{center}
    \includegraphics[height=4.2in]{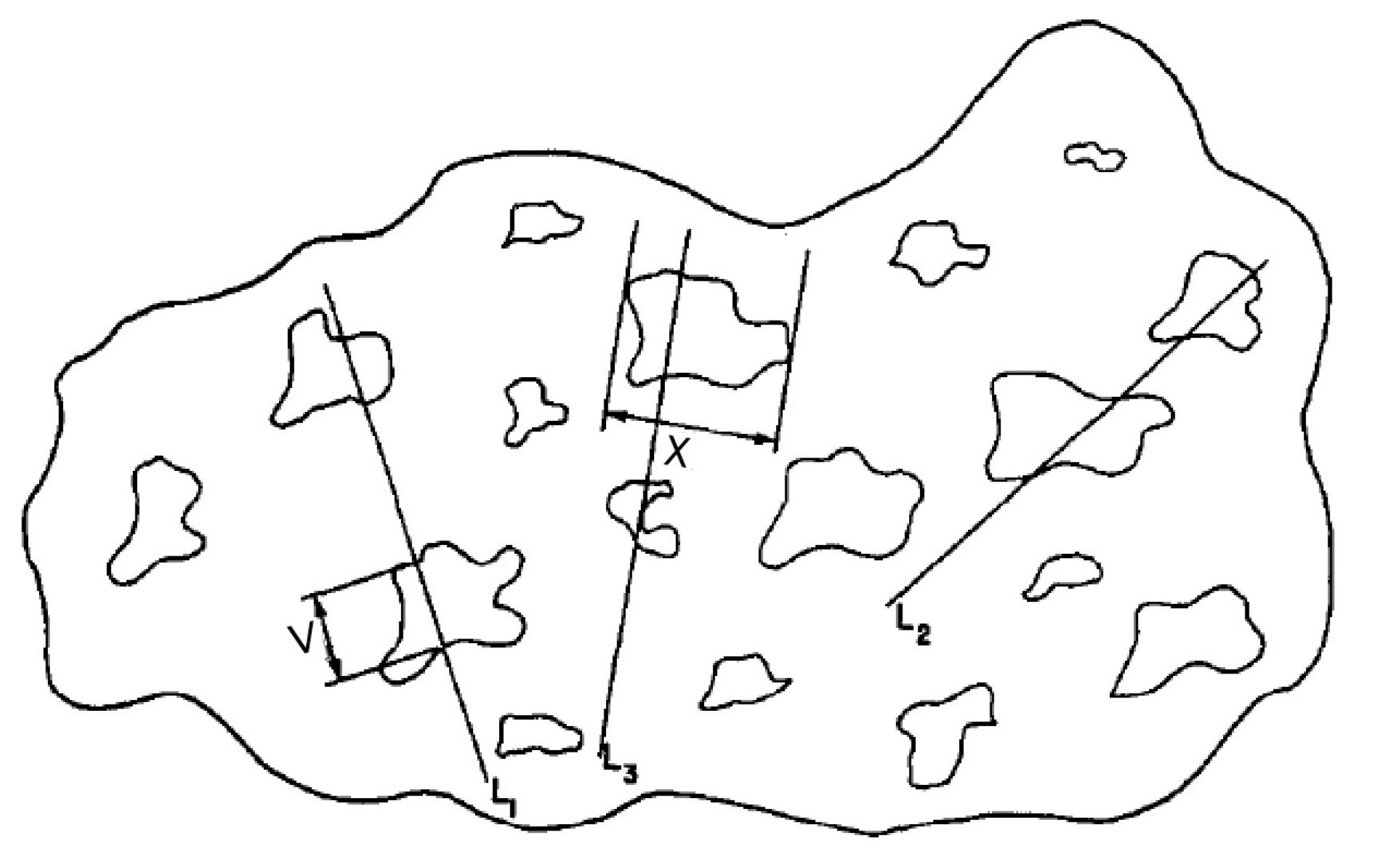}
  \end{center}
  \caption{Sketch of the study area with three transects, $L_1, L_2, L_3$: $X$ is the width of the  intersected 
  shrub perpendicular to the transect and $V$ is the length of intersected shrub parallel to the transect; see Muttlak (1988).}
\end{figure}

\newpage

\begin{figure}[H]
    \centering
    \includegraphics[width=\textwidth, height=8in]{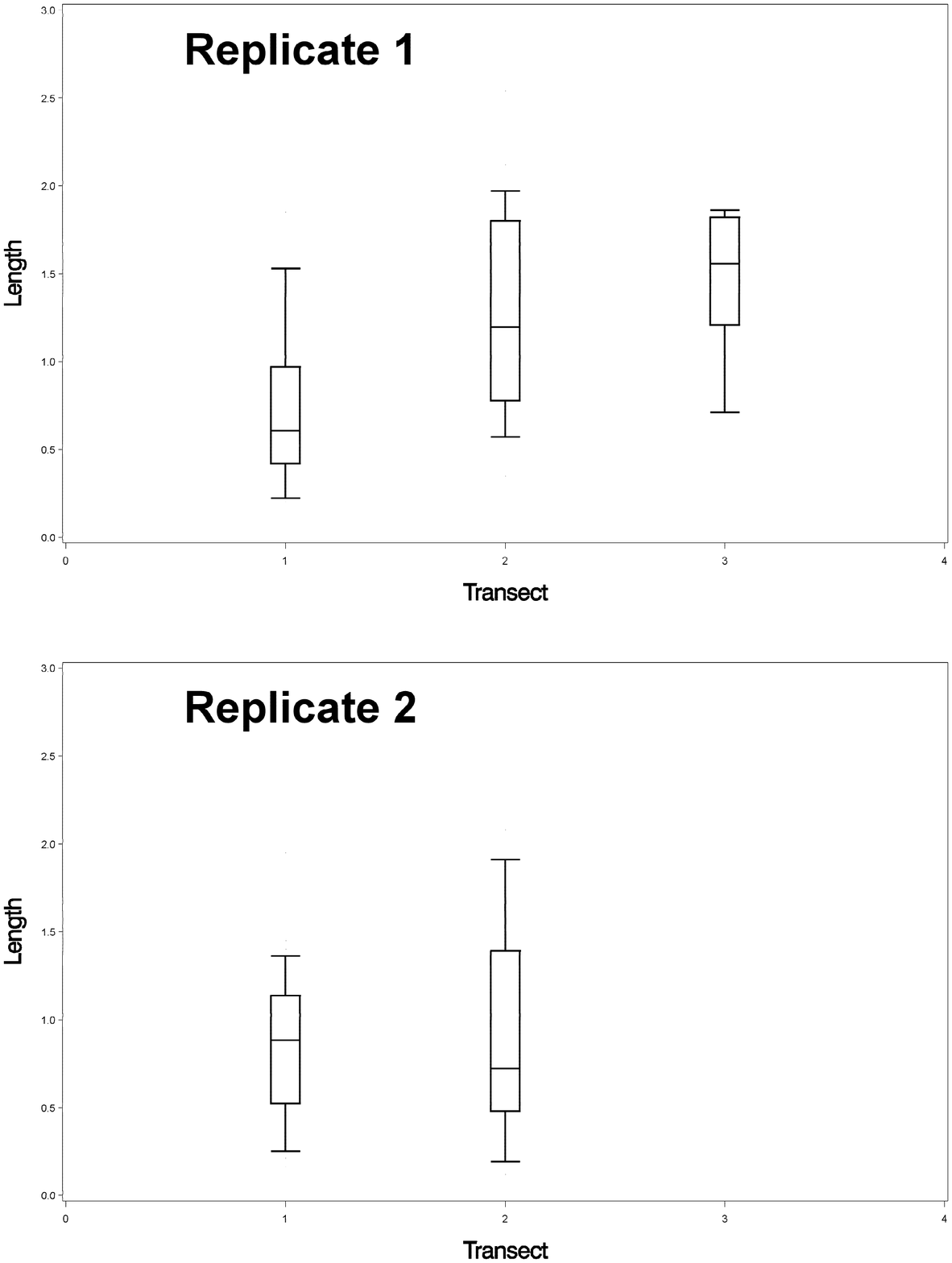}  
    \begin{center}
    \caption{Box plots of the length-biased data from the two replicates} 
    \end{center}
\end{figure}  

%
%\begin{figure}[!htpb]\begin{center}
%\includegraphics[width=3.5in]{box1.pdf}\vspace{2cm}
%\includegraphics[width=3.5in]{box2.pdf}
%\caption{Box plots of Replication 1 (top)  and Replication 2 (bottom)}
%\end{center}\end{figure}

%
%\newpage
%
%\begin{figure}[!htpb]\begin{center}
%\includegraphics[width=2.5in]{xs.pdf}\qquad
%\includegraphics[width=2.5in]{als.pdf}\vspace{1cm}
%\includegraphics[width=2.5in]{be1s.pdf}\qquad
%\includegraphics[width=2.5in]{be2s.pdf}\vspace{1cm}
%\includegraphics[width=2.5in]{be3s.pdf}\qquad
%\includegraphics[width=2.5in]{gas.pdf}\\
%\caption{Posterior distributions of (a) population mean; (b) $\alpha$; (c) $\beta_1$; (d) $\beta_2$; (e) $\beta_3$; (f) $\gamma$}
%\end{center}\end{figure}

\newpage
\begin{figure}[H]
\centering
\subfigure{
\includegraphics[width=.4\textwidth]{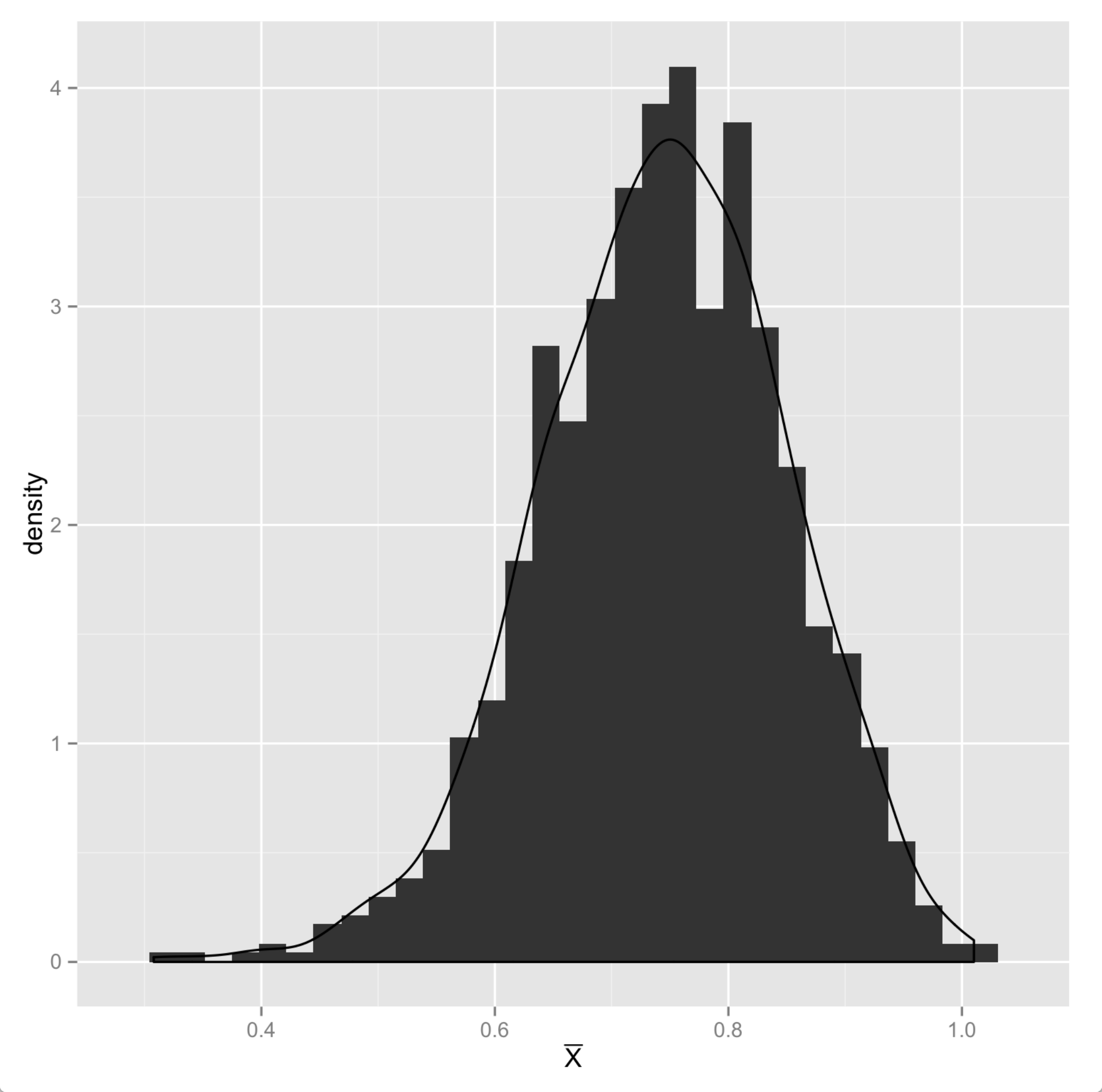}
}
\subfigure{
\includegraphics[width=.4\textwidth]{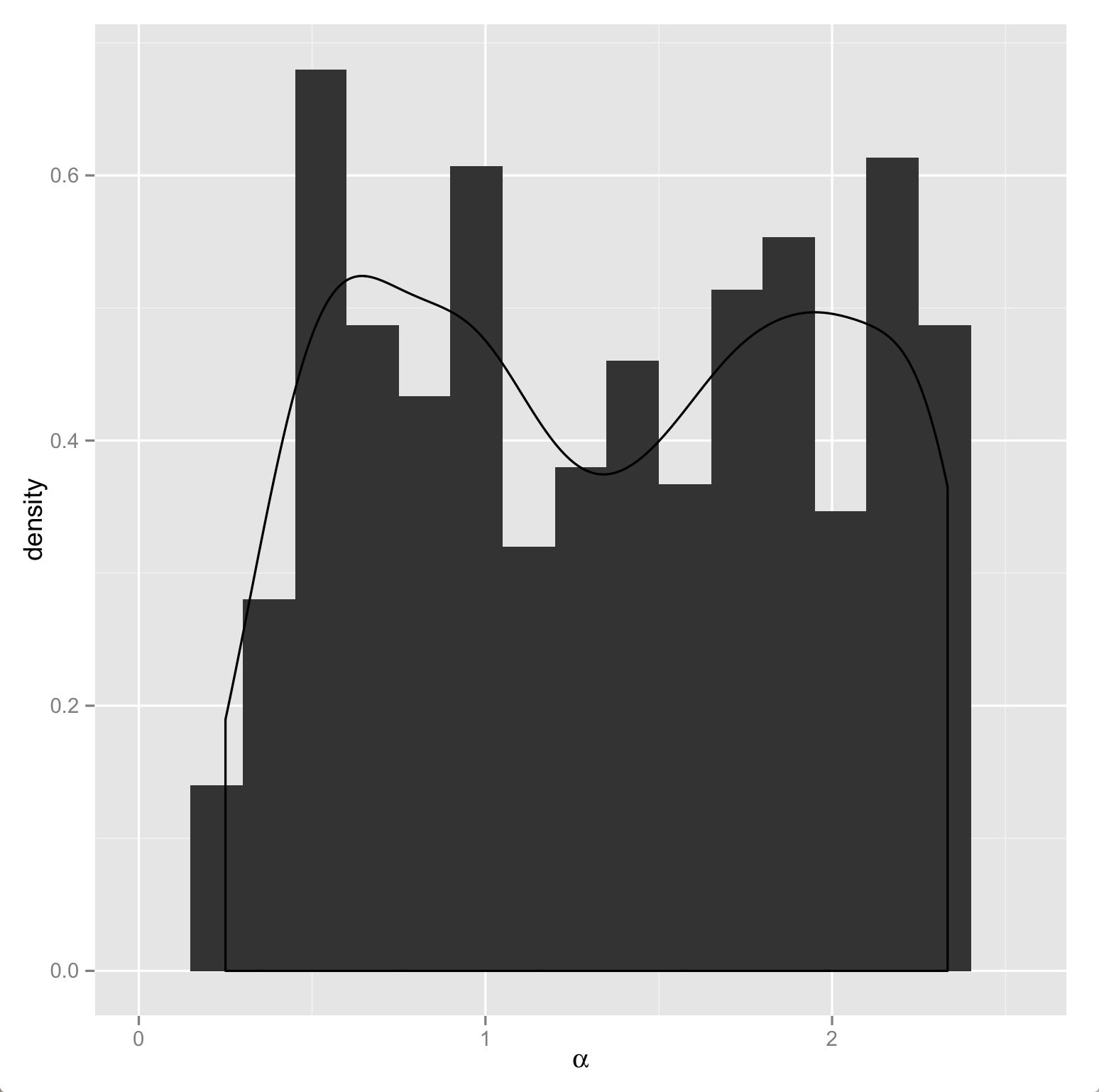}
}

\subfigure{
\includegraphics[width=.4\textwidth]{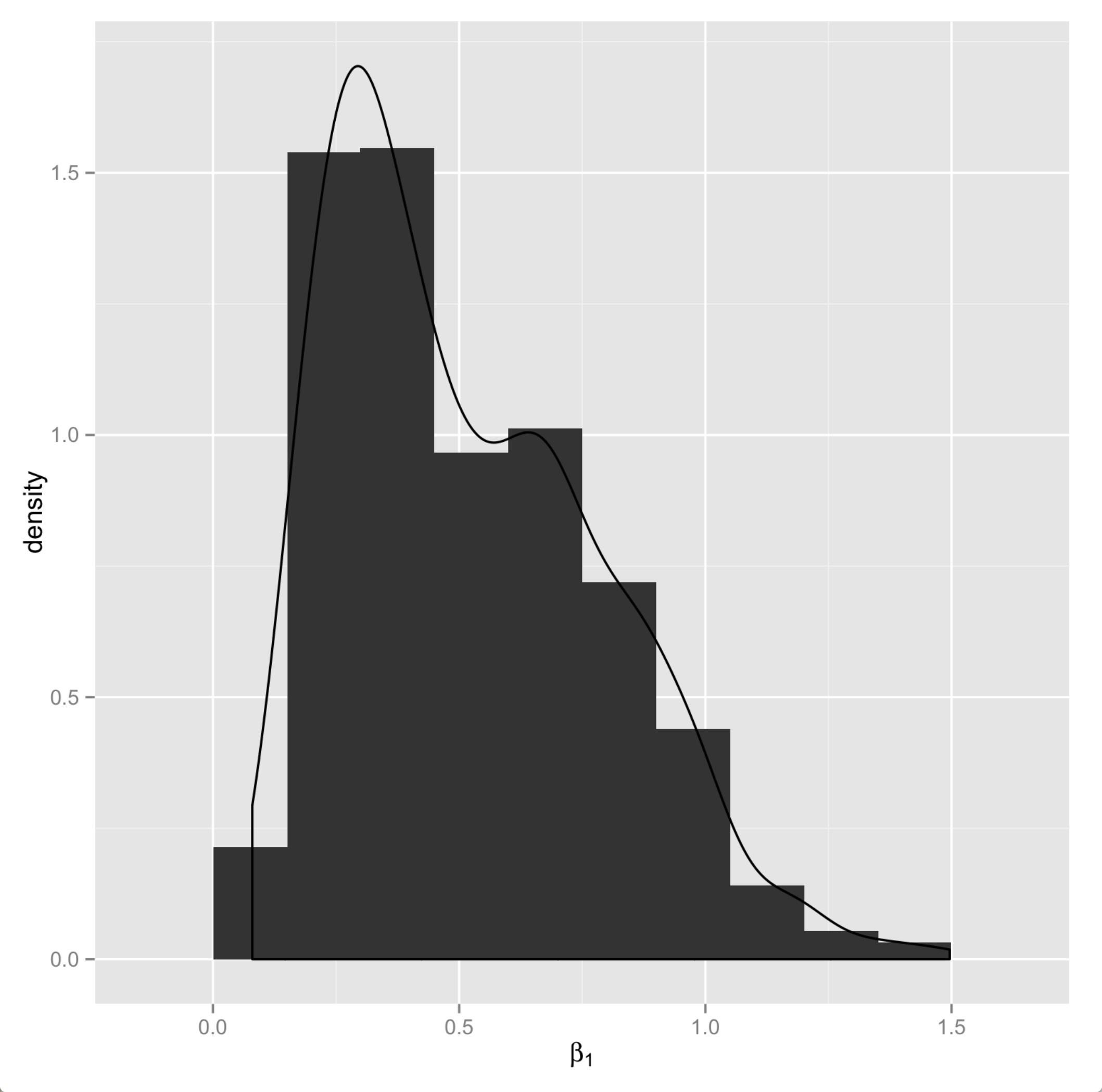}
}
\subfigure{
\includegraphics[width=.4\textwidth]{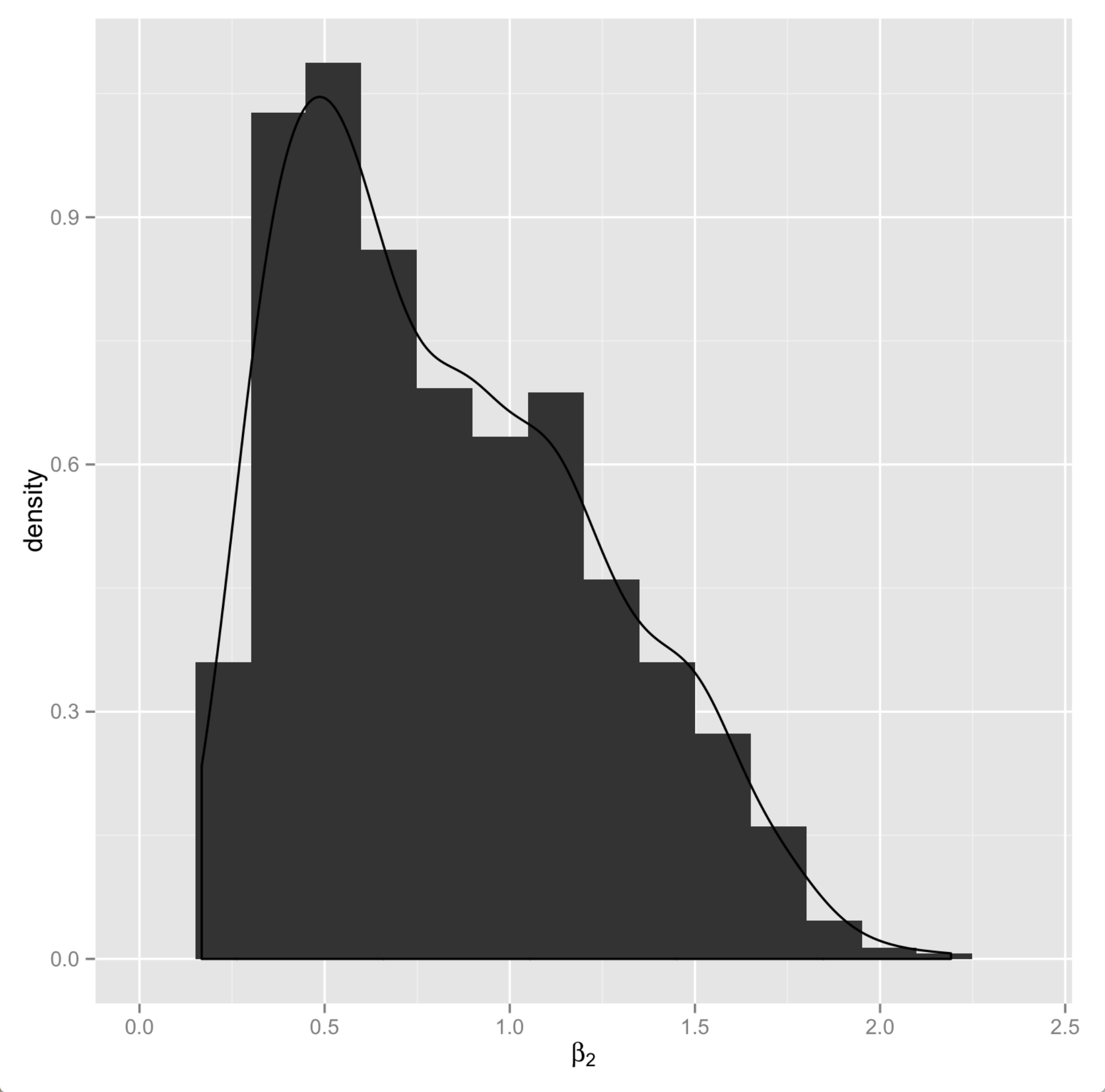}
}

\subfigure{
\includegraphics[width=.4\textwidth]{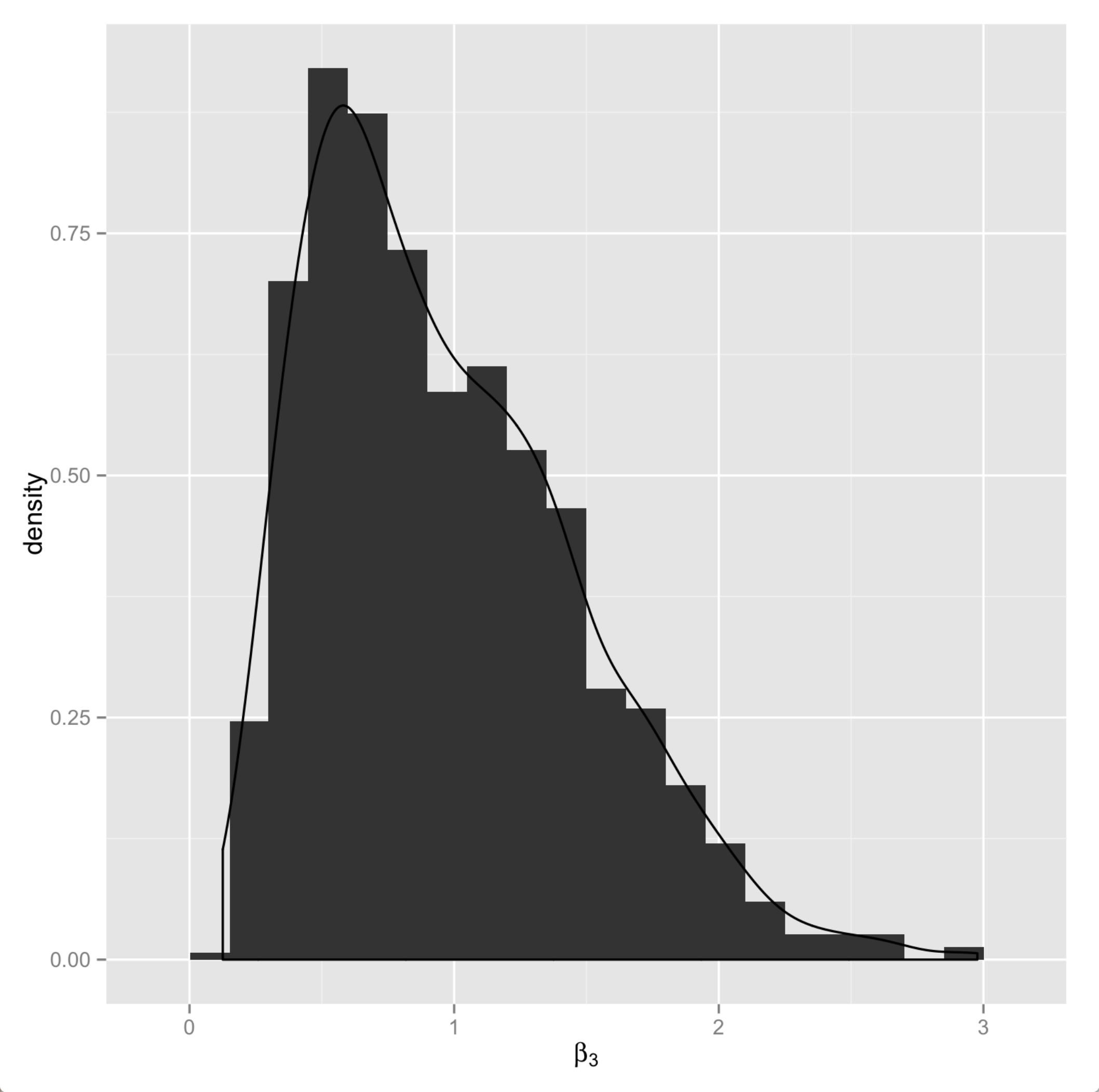}
}
\subfigure{
\includegraphics[width=.4\textwidth]{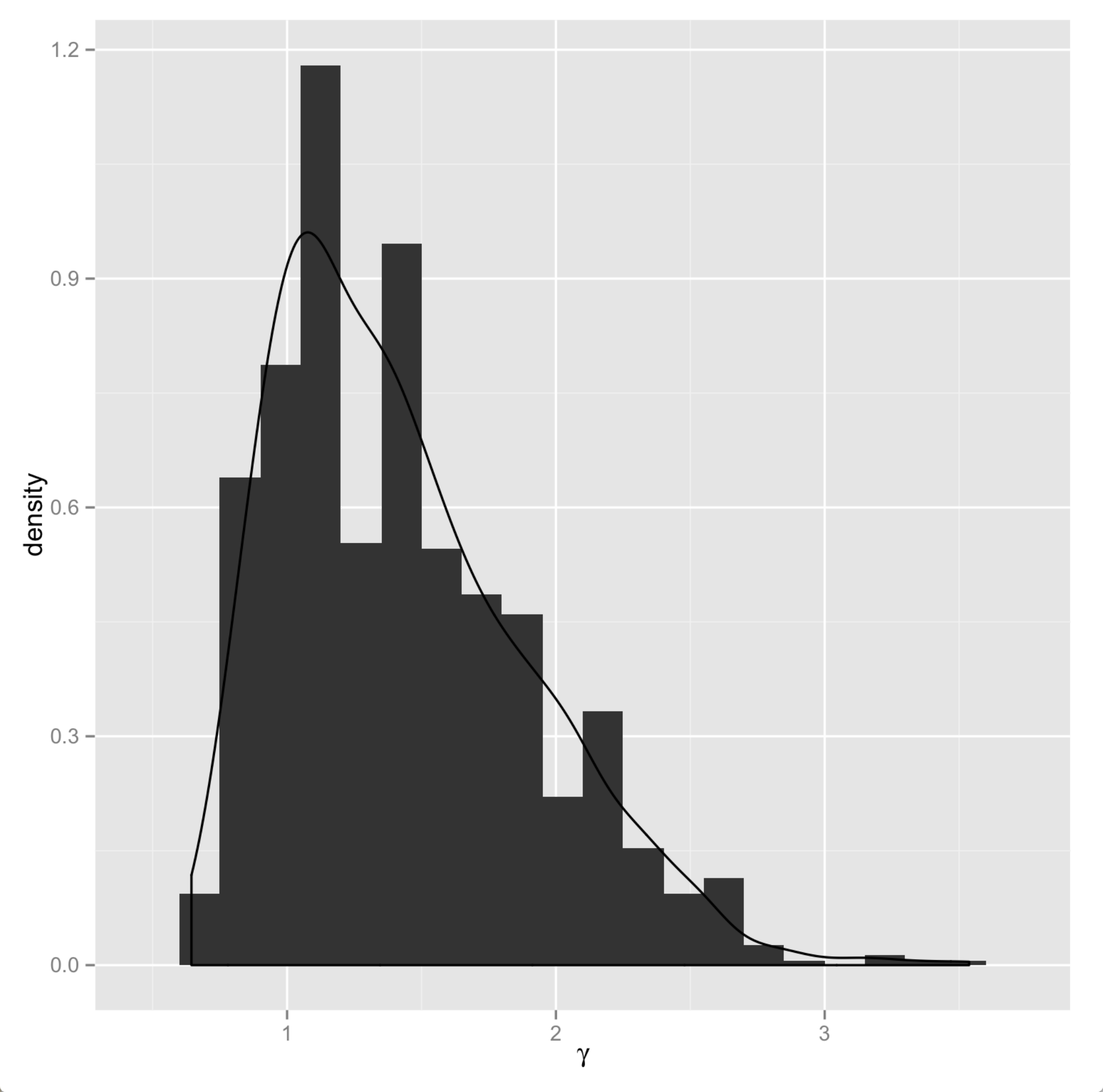}
}
\caption{Posterior distributions of population mean;  $\alpha$; $\beta_1$; $\beta_2$; $\beta_3$;  and $\gamma$}
\end{figure}

\newpage
\begin{figure}[H]
    \centering
    \includegraphics[width=\textwidth, height=8in]{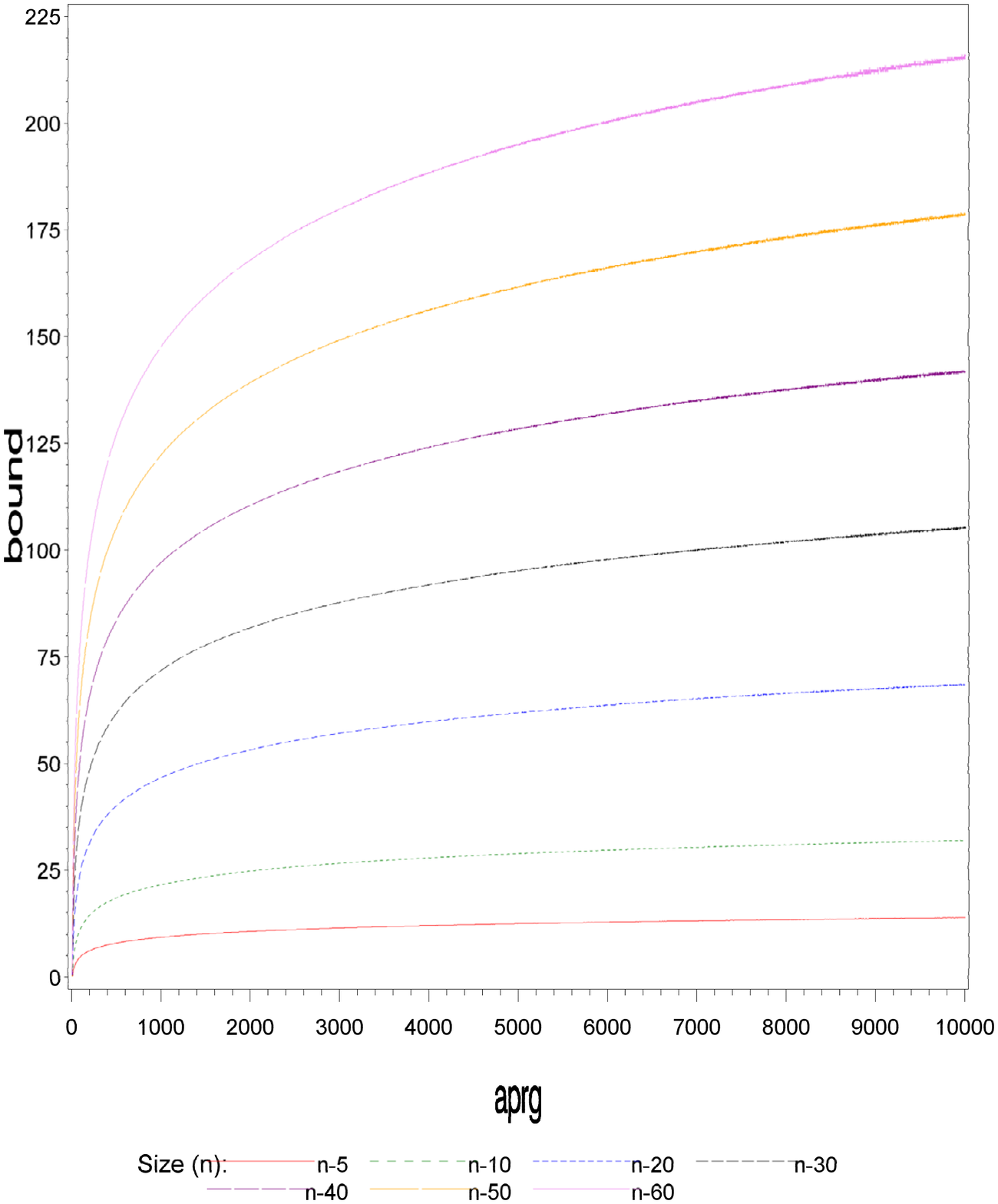}  
    \begin{center}
    \caption{Line plots of $\Delta(\theta)$ for selected sample sizes (n)} 
    \end{center}
\end{figure}   
\newpage
\begin{figure}[H]
    \centering
    \includegraphics[width=\textwidth, height=8in]{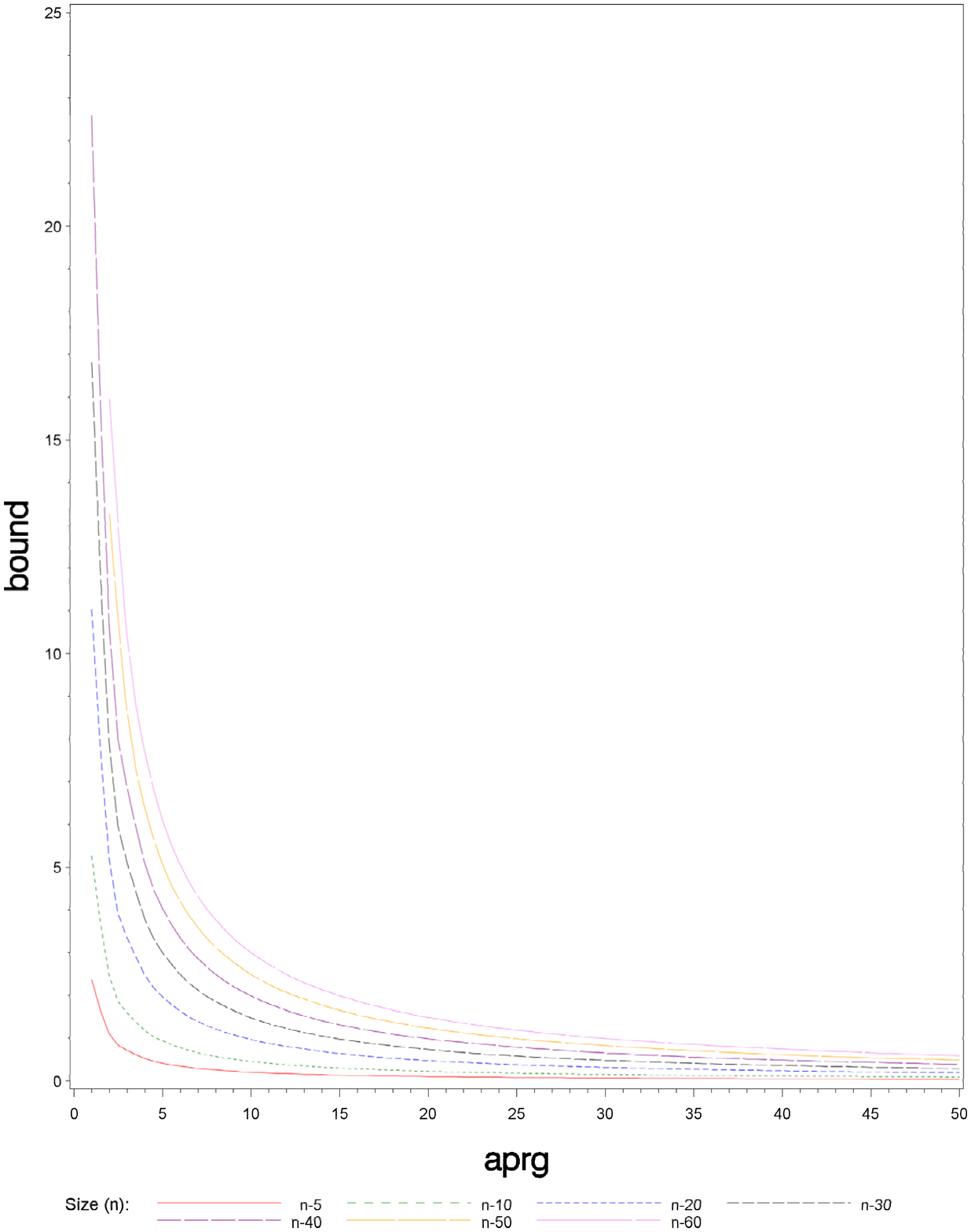}  
    \begin{center}
    \caption{Line plots of $\Delta^\prime(\theta)$ for selected sample sizes (n)} 
    \end{center}
\end{figure}

% \ convert als.png als.pdf
  %\ convert f4.png f4.jpg
   %\ convert box1.png box1.jpg
 %\ convert box2.png box2.jpg

\end{document}